\documentclass[twocolumn,twocolappendix]{aastex631}

\usepackage{xspace}
\usepackage{subfigure}
\usepackage{multirow}

\usepackage{savesym}
\savesymbol{tablenum}
\usepackage{siunitx}
\restoresymbol{SIX}{tablenum}

\newcommand{\gam}{$\gamma$\xspace}
\newcommand{\hess}{H.E.S.S.\xspace}
\newcommand{\tdorc}{30\,Dor\,C\xspace}
\newcommand{\mcsnr}{MCSNR~J0536$-$6913\xspace}
\newcommand{\hessjpwn}{HESS~J0537$-$691\xspace}
\newcommand{\hessjsb}{HESS~J0535$-$691\xspace}
\newcommand{\hessjrmc}{HESS~J0538$-$691\xspace}
\newcommand{\pks}{PKS~2155$-$304\xspace}
\newcommand{\gammapy}{\textsc{Gammapy}\xspace}
\newcommand{\sigmaext}{\ensuremath{\sigma_\mathrm{Gauss}}}

\newcommand{\estat}{\ensuremath{_\mathrm{stat}}}
\newcommand{\esys}{\ensuremath{_\mathrm{sys}}}
\newcommand{\siee}[4]{\ensuremath{(#1\pm #2\estat\pm #3\esys)\,\mathrm{#4}}\xspace}
\newcommand{\hms}[3]{$#1^\mathrm{h}#2^\mathrm{m}#3^\mathrm{s}$\xspace}
\newcommand{\hmse}[4]{$#1^\mathrm{h}#2^\mathrm{m}#3^\mathrm{s}\pm #4^\mathrm{s}\estat$\xspace}

\DeclareSIUnit\pc{pc}
\DeclareSIUnit\kpc{kpc}
\DeclareSIUnit\erg{erg}  
\DeclareSIUnit\year{yr}
\DeclareSIUnit\TeV{TeV}
\DeclareSIUnit\GeV{GeV}
\DeclareSIUnit\PeV{PeV}
\DeclareSIUnit\gauss{G}
\DeclareSIUnit\mas{mas}

\shorttitle{VHE \gam-ray emission from young massive star clusters in the Large Magellanic Cloud}
\shortauthors{F.\ Aharonian et al.}

\graphicspath{{./}{img/}}

\begin{document}

\title{Very-high-energy \gam-ray emission from young massive star clusters in the Large Magellanic Cloud}

\author{F.~Aharonian}
\affiliation{Dublin Institute for Advanced Studies, 31 Fitzwilliam Place, Dublin 2, Ireland}
\affiliation{Max-Planck-Institut f\"ur Kernphysik, Saupfercheckweg 1, 69117 Heidelberg, Germany}
\affiliation{Yerevan State University,  1 Alek Manukyan St, Yerevan 0025, Armenia}

\author{F.~Ait~Benkhali}
\affiliation{Landessternwarte, Universit\"at Heidelberg, K\"onigstuhl, D 69117 Heidelberg, Germany}

\author{J.~Aschersleben}
\affiliation{Kapteyn Astronomical Institute, University of Groningen, Landleven 12, 9747 AD Groningen, The Netherlands}

\author[0000-0002-2153-1818]{H.~Ashkar}
\affiliation{Laboratoire Leprince-Ringuet, \'Ecole Polytechnique, CNRS, Institut Polytechnique de Paris, F-91128 Palaiseau, France}

\author[0000-0002-9326-6400]{M.~Backes}
\affiliation{University of Namibia, Department of Physics, Private Bag 13301, Windhoek 10005, Namibia}
\affiliation{Centre for Space Research, North-West University, Potchefstroom 2520, South Africa}

\author[0000-0002-5085-8828]{V.~Barbosa~Martins}
\affiliation{Deutsches Elektronen-Synchrotron DESY, Platanenallee 6, 15738 Zeuthen, Germany}

\author[0000-0002-5797-3386]{R.~Batzofin}
\affiliation{Institut f\"ur Physik und Astronomie, Universit\"at Potsdam,  Karl-Liebknecht-Strasse 24/25, D 14476 Potsdam, Germany}

\author[0000-0002-2115-2930]{Y.~Becherini}
\affiliation{Universit\'e de Paris, CNRS, Astroparticule et Cosmologie, F-75013 Paris, France}
\affiliation{Department of Physics and Electrical Engineering, Linnaeus University,  351 95 V\"axj\"o, Sweden}

\author[0000-0002-2918-1824]{D.~Berge}
\affiliation{Deutsches Elektronen-Synchrotron DESY, Platanenallee 6, 15738 Zeuthen, Germany}
\affiliation{Institut f\"ur Physik, Humboldt-Universit\"at zu Berlin, Newtonstr. 15, D 12489 Berlin, Germany}

\author[0000-0001-8065-3252]{K.~Bernl\"ohr}
\affiliation{Max-Planck-Institut f\"ur Kernphysik, Saupfercheckweg 1, 69117 Heidelberg, Germany}

% ...

\collaboration{300}{(\hess Collaboration)}

\NewPageAfterKeywords

\begin{abstract}
The Tarantula Nebula in the Large Magellanic Cloud is known for its high star formation activity.
At its center lies the young massive star cluster R136, providing a significant amount of the energy that makes the nebula shine so brightly at many wavelengths.
Recently, young massive star clusters have been suggested to also efficiently produce high-energy cosmic rays, potentially beyond PeV energies.
Here, we report the detection of very-high-energy \gam-ray emission from the direction of R136 with the High Energy Stereoscopic System, achieved through a multicomponent, likelihood-based modeling of the data.
This supports the hypothesis that R136 is indeed a very powerful cosmic-ray accelerator.
Moreover, from the same analysis, we provide an updated measurement of the \gam-ray emission from \tdorc, the only superbubble detected at TeV energies presently.
The \gam-ray luminosity above \SI{0.5}{TeV} of both sources is $(2-3)\times 10^{35}\,\mathrm{erg}\,\mathrm{s}^{-1}$.
This exceeds by more than a factor of 2 the luminosity of HESS~J1646$-$458, which is associated with the most massive young star cluster in the Milky Way, Westerlund~1.
Furthermore, the \gam-ray emission from each source is extended with a significance of $>3\sigma$ and a Gaussian width of about \SI{30}{pc}.
For \tdorc, a connection between the \gam-ray emission and the nonthermal X-ray emission appears likely.
Different interpretations of the \gam-ray signal from R136 are discussed.
\end{abstract}


\section{Introduction} \label{sec:intro}
It has been known for many decades that cosmic rays (CRs) with extremely high energies reach us on Earth \citep{PDG2023}.
In recent years, observations of \gam-rays with PeV energies from throughout the Galaxy have confirmed the long-standing hypothesis that CRs with multi-PeV energies are produced within the Milky Way \citep{TibetASg2021,LHAASO2023}.
Despite decades of searches, their precise origins are, however, still unresolved.
While shock fronts of young supernova remnants (SNRs) have long been considered as the main acceleration sites of CR nuclei \citep[``hadronic CRs''; e.g.][]{Ginzburg1964,Berezinskii1990}, the potential of stellar winds to accelerate CRs was also realized early on \citep[e.g.][]{Cesarsky1983}.
In the last few years, young massive star clusters (YMCs) have increasingly been discussed as potentially predominant sources of the highest-energy Galactic CRs \citep[e.g.][]{Aharonian2019,Morlino2021,Vieu2023}.
If YMCs generate high-energy hadronic CRs, they are expected to also be sources of \gam-rays, which are created predominantly in the decay of neutral pions that emerge when the CRs interact with ambient gas.
This is referred to as the ``hadronic scenario'' for the generation of high-energy \gam-ray emission.
The hypothesis that YMCs are effective CR accelerators can therefore be tested through observations in the very-high-energy (VHE; $E>\SI{0.1}{TeV}$) \gam-ray domain.

In fact, \citet{HESS_Wd1_2022} have recently been able to associate the VHE \gam-ray source HESS~J1646$-$458 with Westerlund~1, the most massive young star cluster in our Galaxy, thus revealing it as a powerful particle accelerator.
This does not yet constitute unequivocal evidence for the acceleration of hadronic CRs by the cluster, however, since the nature of the emitting particles remains ambiguous.
In fact, for the case of Westerlund~1, \citet{Haerer2023} have demonstrated that the morphology is inconsistent with the standard hadronic scenario, and that a model that explains the \gam-ray emission as being due to inverse-Compton (IC) scattering of CR \emph{electrons} (the ``leptonic scenario'') provides a more natural explanation of the High Energy Stereoscopic System (\hess) measurements.
Moreover, the exact acceleration site remains unidentified; proposals in the literature include shocks forming at the interaction of winds of massive stars inside the cluster \citep[e.g.][]{Bykov2013}, the termination shock of the collective cluster wind \citep{Gupta2020,Morlino2021}, and magnetic turbulences within the entire superbubble (SB) blown by the cluster wind \citep[e.g.][and references therein]{Vieu2022}.
A definitive observational confirmation of any of these predictions is still lacking.
Unfortunately, only about a handful of YMCs in the Milky Way have been detected in the VHE domain so far \citep[see, e.g.,][]{HESS_Wd1_2022}, and the association of the \gam-ray emission with the star cluster is not always firm.
The detection and observation of further YMCs with VHE \gam-rays could therefore help to shed more light on their role as CR accelerators.

The Large Magellanic Cloud (LMC), containing many massive star clusters, is a promising target to search for \gam-ray emission from YMCs.
Indeed, it is host to \tdorc, an SB inflated by the LH~90 association of star clusters \citep{Lucke1970,Lortet1984} that is visible not only in the radio and optical domains \citep{Mathewson1985} but also in nonthermal X-rays \citep{Bamba2004,Smith2004,Yamaguchi2009,Kavanagh2015,Kavanagh2019,Lopez2020}, indicating the presence of high-energy electrons.
\tdorc is the only confirmed SB that has been detected in VHE \gam-rays so far\footnote{The \gam-ray emission from Westerlund~1 is very likely also associated with the SB of that cluster. In that case, however, a firm detection of the SB at other wavelengths is lacking.} \citep{HESS_LMC_2015}.
Located nearby, at the heart of the Tarantula Nebula and its central open cluster NGC~2070, lies the ``super star cluster'' R136, which is exceptionally rich in massive stars \citep{Crowther2010}.
With an estimated age between \SIrange{1}{2}{Myr} \citep[e.g.][]{Massey1998,Brands2022}, R136 is also relatively young, implying that only few supernovae (SNe) are expected to have occurred since its birth (although some older massive stars have also been found in NGC~2070, see \citealt{Sabbi2012,Schneider2018,Bestenlehner2020}).

In this paper, we report the discovery of VHE \gam-ray emission from the direction of R136 with the High Energy Stereoscopic System (\hess), achieved through a multicomponent, likelihood-based modeling of the spatial and spectral distribution of \gam-ray-like events.
Moreover, from the same analysis, we provide updated results on the SB \tdorc.
Throughout the paper, we assume a distance to the LMC of \SI{50}{kpc} \citep{Pietrzynski2013}.

\section{Data Analysis}\label{sec:data_analysis}
\hess is a \gam-ray observatory located in the Khomas highland of Namibia, at \SI{1800}{m} above sea level.
In its initial configuration that began operating in 2003, it consisted of four identical Cherenkov telescopes with \SI{107}{\square\meter} mirror area each, arranged in a square with \SI{120}{m} side length \citep{HESS_Crab_2006}.
In this configuration, \hess is sensitive to \gam-rays above a few hundred GeV.
In 2012, the array was augmented with a fifth, larger telescope with \SI{600}{\square\meter} mirror area, extending the sensitivity range to energies below \SI{100}{GeV}.

\hess has collected an extensive data set on the Tarantula Nebula region.
Observations are taken in individual ``runs'' of usually \SI{28}{\minute} duration.
Here, we have analyzed a total of 794 runs taken with all four of the initial telescopes, corresponding to an observation time of $\approx$\SI{360}{\hour}.
Of these, 301~runs ($\approx$\SI{138}{\hour}) have been taken between 2004 December 30 and 2013 February 7, while the remaining 493~runs ($\approx$\SI{222}{\hour}) date to between 2017 October 14 and 2022 February 2.\footnote{The reduced exposure time of the pre-2014 data set with respect to \citet{HESS_LMC_2015} is due to stricter selection criteria, in particular due to the requirement that all four \SI{107}{\square\meter} telescopes participate in each run. This requirement also explains the omission of observations taken between 2013 and 2017, most of which have been taken with an incomplete array, largely due to an upgrade of the cameras of the initial telescopes that took place during this time \citep{Ashton2020}.}
Because a substantial fraction of the observations have been carried out prior to the installation of the fifth telescope, only uniform data from the four initial telescopes are being considered.

The detection of \gam-rays proceeds via measuring the Cherenkov radiation that is emitted by secondary particles in the extensive air shower that is launched when the primary \gam-ray impinges on the atmosphere of the Earth.
We employ the \textsc{ImPACT} algorithm \citep{Parsons2014} to reconstruct the incoming direction and energy of the primary particles (``events'') from the telescope camera images.
Instead of the customary two telescopes, we require that every event is detected by at least three telescopes, which enhances the angular resolution of the instrument at the expense of a slight reduction in effective area.
For optimal performance, we furthermore reject events with a reconstructed direction that deviates from the pointing direction of the telescopes by more than $1.5^\circ$.
Owing to the relatively strict selection cuts, the achieved energy threshold for the data set is \SI{0.5}{TeV}.
Instrument response functions (IRFs) have been generated from extensive Monte Carlo simulations \citep{Bernloehr2008}.
The suppression of ``hadronic'' background events, which consist primarily of air showers initiated by CR nuclei, is performed using the method described in \citet{Ohm2009}.
We perform high-level analysis of the data with \gammapy~v0.18.2 \citep{Donath2023,Deil2020}, employing a three-dimensional binned likelihood fit \citep{Mattox1996} in a $5^\circ\times 5^\circ$ region of interest (ROI), with spatial pixels of size $0.02^\circ\times 0.02^\circ$ (see Table~\ref{tab:analysis_settings} in Appendix~\ref{sec:appx_bkg_fit} for details).
In the analysis, the residual level of background in the final event sample is described using a model constructed from archival \hess observations, following the procedure outlined in \citet{Mohrmann2019}.
We demonstrate in Appendix~\ref{sec:appx_bkg_fit} that after an adjustment of the background model to each observation run, we obtain a good description of the residual hadronic background for the full data set.

After the adjustment of the background model, we proceed with modeling of the \gam-ray emission, following an iterative procedure in which we continue to add sources to the ROI model until no significant \gam-ray emission remains.
This procedure is described in more detail in Appendix~\ref{sec:appx_src_fit}.
As a result, we obtain a best-fit spatial and spectral model for each source included in the model.
As spatial models, we use two-dimensional, radially symmetric Gaussians with variable width $\sigmaext$, while the energy spectra are modeled with either a power law or a log-parabola function.
Furthermore, we employ the \textsc{naima} package \citep{Zabalza2015} to fit physical spectral models of primary CR particles to our data.
All modeling results have been cross-checked using a second analysis pipeline that is based on an independent calibration, event reconstruction, and event selection \citep{deNaurois2009}.

\begin{figure*}[ht]
  \centering
  \includegraphics{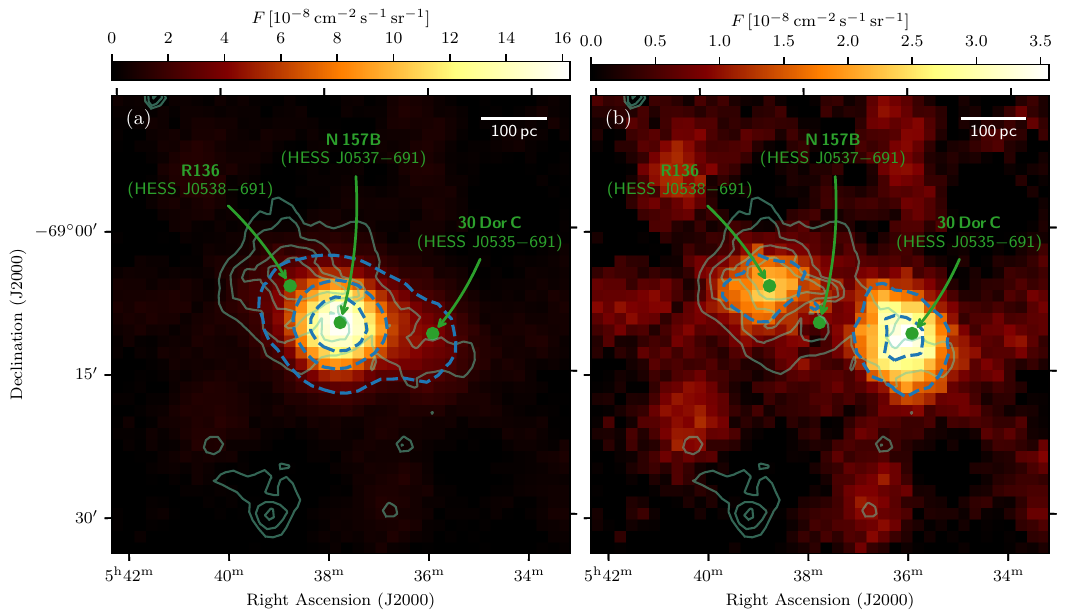}
  \caption{
    \gam-ray flux maps of the target region of the analysis.
    The maps show the \gam-ray flux $F$ in units of $10^{-8}\,\mathrm{cm}^{-2}\,\mathrm{s}^{-1}\,\mathrm{sr}^{-1}$, integrated above an energy of \SI{0.5}{TeV}, assuming a power-law spectrum with index $-2.5$.
    Smoothing with a top-hat kernel of $0.07^\circ$ radius has been applied.
    (a) Entire emission.
    (b) Residual emission after subtraction of the emission from N\,157B predicted by the best-fit ROI model.
    Dashed blue lines show flux contours at $(2.5/7.5/12.5)\times 10^{-8}\,\mathrm{cm}^{-2}\,\mathrm{s}^{-1}\,\mathrm{sr}^{-1}$ in panel \emph{a} and $(1.5/3)\times 10^{-8}\,\mathrm{cm}^{-2}\,\mathrm{s}^{-1}\,\mathrm{sr}^{-1}$ in panel \emph{b}.
    Pixels with a negative excess after background subtraction are clipped at zero.
    The light green contour lines denote H$\alpha$ emission as inferred by SHASSA \citep{Gaustad2001}.
  }
  \label{fig:flux_maps}
\end{figure*}

\section{Results}\label{sec:results}
We show a \gam-ray flux map of the Tarantula Nebula region in Fig.~\ref{fig:flux_maps}(a).
The applied smoothing is representative of the \hess angular resolution for the employed analysis configuration.
The nebula is outlined by the H$\alpha$ emission contours from the Southern H-Alpha Sky Survey Atlas \citep[SHASSA;][]{Gaustad2001}.
As can be seen on the map, by far the brightest \gam-ray source in this region is the pulsar wind nebula (PWN) N\,157B, also known as \hessjpwn, which is associated with the pulsar PSR~J0537$-$6910 \citep{HESS_N157B_2012,HESS_LMC_2015}.
To test for the presence of additional sources, we performed a spectromorphological modeling of the \gam-ray emission, adding source components -- beginning with a model for N~157B -- to the ROI model until no significant residual emission remains.
The modeling, described in detail in Appendix~\ref{sec:appx_src_fit}, reveals the presence of two additional sources: \hessjsb, previously associated with \tdorc\ \citep{HESS_LMC_2015}, and a new source, \hessjrmc, whose association with R136 we argue for in this paper.
This is illustrated in Fig.~\ref{fig:flux_maps}(b), where the emission from N\,157B as predicted by our best-fit ROI model has been subtracted, thus making evident the remaining emission from the two other sources.
Best-fit parameter values for all modeled sources can be found in Appendix~\ref{sec:appx_src_fit}, as can an explanation of the method we use to derive systematic uncertainties on these values.
In what follows, we highlight the most relevant results.

\subsection{Description of source models}
\emph{N~157B} --- We do not investigate N\,157B in detail in this paper, but note that our analysis yields a nonzero extension of $\sigmaext^\mathrm{N\,157B}=(0.82\pm 0.20\estat\pm 0.18\esys)'$ for this source.
We caution, however, that the extended source model is preferred over a pointlike model by only $\approx$1.3$\sigma$ -- these seemingly contradictory results are due to the presence of the other two nearby sources, whose model components can absorb part of the emission for the case that N\,157B is less extended than suggested by our best fit.
We therefore do not claim an extension and provide an upper limit of $\sigmaext^\mathrm{N\,157B}<1.14'$ (95\% confidence level, statistical uncertainties only).
Whether extended or not, as we demonstrate in Appendix~\ref{sec:appx_src_fit}, the results obtained for \tdorc and R136 do not depend strongly on the model assumed for N~157B.
The obtained spectrum for N~157B, shown in Fig.~\ref{fig:sed_n157b} in Appendix~\ref{sec:appx_src_fit}, is compatible with our previously published result.

\begin{figure}
  \centering
  \includegraphics{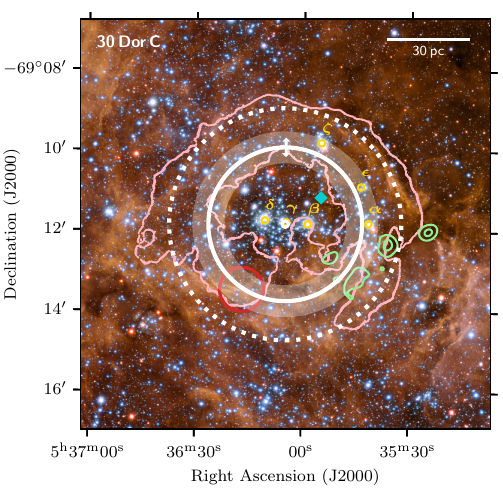}
  \caption{
    Optical image (credit: ESO; \url{https://www.eso.org/public/images/eso1816a}) showing the LH~90 association of star clusters.
    The yellow circles mark the positions of the compact clusters $\alpha$--$\zeta$ found by \citet{Lortet1984}.
    The pink contour line outlines the SB \tdorc as visible in X-rays (\SIrange{1}{2}{keV}) with \emph{XMM-Newton}; the red circle denotes \mcsnr, a putative SNR \citep{Kavanagh2015}.
    The white circle marker, solid line, and dotted line indicate the best-fit position, 1-$\sigma$ Gaussian radius, and 68\% containment radius of the best-fit \gam-ray model for \tdorc, respectively.
    The transparent band shows the statistical uncertainty on the 1-$\sigma$ radius, while the arrows denote the systematic uncertainty.
    The turquoise diamond marks the position reported in \citet{HESS_LMC_2015}.
    The green contour lines denote bright spots of $^{12}$CO(1--0) emission measured with the Atacama Large Millimeter/submillimeter Array \citep{Yamane2021}.
  }
  \label{fig:vlt_optical_image_tdorc}
\end{figure}

\emph{\tdorc} --- For the first time, we find \hessjsb, associated with \tdorc, to be an extended \gam-ray source.
The best-fit Gaussian width is $\sigmaext^\mathrm{\tdorc}=(1.91\pm 0.40\estat\pm 0.20\esys)'$, which corresponds to $\siee{27.8}{5.8}{2.9}{pc}$ at the distance to the LMC.
This model is preferred over one in which \tdorc is described as a pointlike source by 3.3$\sigma$.
The measured extension is of the same order as the observed size of the X-ray SB, as can be seen from Fig.~\ref{fig:vlt_optical_image_tdorc}.
The best-fit position deviates by $1.1'$ from that previously obtained in \citet{HESS_LMC_2015}; this is most likely due to the different analysis method used there (a two-dimensional, i.e.\ energy-integrated likelihood fit).
We note that the new position is in better agreement with the center of the X-ray SB and the compact star clusters located there.
The energy spectrum follows a power law with spectral index $\Gamma_\mathrm{\tdorc}=-2.57\pm 0.09\estat$.

\begin{figure}
  \centering
  \includegraphics{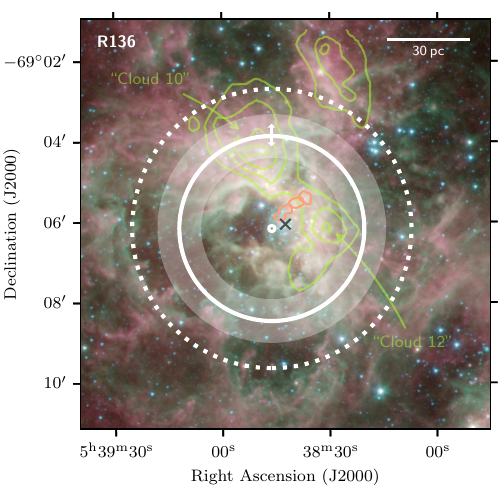}
  \caption{
    Composite infrared image from \emph{Spitzer} (credit: NASA/JPL-Caltech; \url{https://www.spitzer.caltech.edu/image/ssc2020-06b-tarantula-nebula-spitzer-3-color-image}) of the region around the star cluster R136.
    The nominal position of R136 \citep{Hog2000} is marked with a gray cross.
    The white circle marker, solid line, and dotted line indicate the best-fit position, 1-$\sigma$ Gaussian radius, and 68\% containment radius of the best-fit \gam-ray model for R136, respectively.
    The transparent band shows the statistical uncertainty on the 1-$\sigma$ radius, while the arrows denote the systematic uncertainty.
    The green contour lines denote $^{12}$CO(1--0) emission \citep{Johansson1998}, while the light red contours indicate the positions of dense ``knots'' of molecular gas identified through $^{12}$CO(2--1) emission \citep{Kalari2018}.
  }
  \label{fig:spitzer_ir_image_r136}
\end{figure}

\emph{R136} --- Lastly, \hessjrmc is detected as a new \gam-ray source with a significance of 6.3$\sigma$.
The separation between the best-fit position and the location of the YMC R136 is only $\approx 20''$ (see Fig.~\ref{fig:spitzer_ir_image_r136}).
Because there is no other plausible counterpart, we associate \hessjrmc with R136.\footnote{The position of \hessjrmc is also compatible with a nearby ``clump'' of stars \citep[separated by a few parsecs;][]{Sabbi2012} that do not belong to R136 itself but are part of the encompassing cluster NGC~2070. See also Sect.~\ref{sec:discussion}.}
Similarly to \tdorc, we find a preference (3.1$\sigma$) for an extended source, with a Gaussian width of $\sigmaext^\mathrm{R136}=(2.30\pm 0.54\estat\,_{-0.21}^{+0.26}|\esys)'$, or $(33.5\pm 7.9\estat\,_{-3.1}^{+3.8}|\esys)\,\mathrm{pc}$.
For the energy spectrum, we find a power-law spectral index $\Gamma_\mathrm{R136}=-2.54\pm 0.15\estat$.

\subsection{Spectral results and energy requirements}
\begin{figure}
  \centering
  \includegraphics{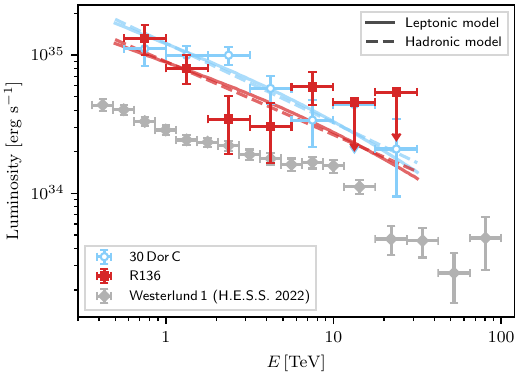}
  \caption{
    \gam-ray luminosity of \tdorc and R136.
    Shown on the vertical axis is the power output per logarithmic energy interval, that is, $E^2\cdot\,(\mathrm{d}N/\mathrm{d}E)\,\cdot\,4\pi d^2$, with $d$ the distance to the source.
    For comparison, the \gam-ray luminosity of the YMC Westerlund~1 is also shown \citep{HESS_Wd1_2022}.
    The solid and dashed lines display a leptonic and a hadronic model, respectively, which we have fitted to both sources using the \textsc{naima} package \citep{Zabalza2015}.
  }
  \label{fig:luminosities}
\end{figure}

In Fig.~\ref{fig:luminosities}, we show the energy spectra of \tdorc and R136 in terms of their \gam-ray luminosity.
Above the threshold energy of \SI{0.5}{TeV}, the integrated luminosities are $L_\gamma^\mathrm{\tdorc}$$\,\approx\,$\SI{2.9e35}{\erg\per\second} and $L_\gamma^\mathrm{R136}$$\,\approx\,$\SI{2.2e35}{\erg\per\second}.
Remarkably, these values exceed the luminosity of Westerlund~1 above the same energy, \SI{8e34}{\erg\per\second} \citep{HESS_Wd1_2022}, by a factor of 2--3.

Fig.~\ref{fig:luminosities} also shows the \gam-ray luminosities predicted by a leptonic and a hadronic model (see Appendix~\ref{sec:appx_naima_results} for details).
For the hadronic case, we find spectral indices of the primary proton spectrum of $\Gamma_\mathrm{p}^\mathrm{\tdorc}$$\,=\,$$-2.64\pm 0.08\estat$ and $\Gamma_\mathrm{p}^\mathrm{R136}$$\,=\,$$-2.59\pm 0.13\estat$.
Assuming an extrapolation of the particle spectrum to \SI{1}{GeV} with the same spectral index, this would imply a total energy requirement for protons of $W_\mathrm{p}^\mathrm{\tdorc}\approx\num{2.1e53}\,(n/\SI{1}{\per\cubic\centi\meter})^{-1}\,\si{erg}$ and $W_\mathrm{p}^\mathrm{R136}\approx\num{1.1e53}\,(n/\SI{1}{\per\cubic\centi\meter})^{-1}\,\si{erg}$, respectively, where $n$ is the average target gas density.
However, at least for \tdorc such an extrapolation would violate the upper limit on the \gam-ray flux in the \SIrange{1}{10}{GeV} range provided by the \emph{Fermi}-LAT instrument \citep{FermiLAT2016}.
To respect the limit, the primary proton spectrum would, for example, need to transition to a harder spectral index of $-2$ below $\sim$\SI{1}{TeV}.
In this case, one obtains a lower requirement of $W_\mathrm{p}^\mathrm{\tdorc}\approx \num{1.4e52}\,(n/\SI{1}{\per\cubic\centi\meter})^{-1}\,\si{erg}$ (or, for comparison, although no limit from \emph{Fermi}-LAT is available, $W_\mathrm{p}^\mathrm{R136}\approx\num{9.7e51}\,(n/\SI{1}{\per\cubic\centi\meter})^{-1}\,\si{erg}$).
These values can be treated as lower limits for the required energy in protons.

For the leptonic model, we include as IC target radiation fields the cosmic microwave background, infrared-to-optical radiation from dust and stars, and ultraviolet radiation specifically from the massive stars in the clusters themselves (see Appendix~\ref{sec:appx_ic_targets} for more details).
We obtain primary electron spectral indices of $\Gamma_\mathrm{e}^\mathrm{\tdorc}=-3.27\pm 0.11\estat$ and $\Gamma_\mathrm{e}^\mathrm{R136}=-3.19\pm 0.17\estat$.
Given the age of the star clusters and the \hess energy range, this represents a population of cooled electrons, with spectra steepened compared to injection spectra.
Assuming that the primary spectra extend down to at least \SI{0.1}{TeV}, and taking into account the energy-dependent cooling due to IC scattering, we find a minimum required injection power for electrons of $L_\mathrm{e}^\mathrm{\tdorc}\approx\SI{6.4e36}{\erg\per\second}$ for \tdorc and $L_\mathrm{e}^\mathrm{R136}\approx\SI{3.9e36}{\erg\per\second}$ for R136.
In the presence of a magnetic field of $B=\SI{5}{\micro\gauss}$, a value roughly representative for the LMC as a whole \citep{Gaensler2005}, additional synchrotron losses lead to larger requirements of $L_\mathrm{e}^\mathrm{\tdorc}\approx\SI{8.5e36}{\erg\per\second}$ and $L_\mathrm{e}^\mathrm{R136}\approx\SI{5.3e36}{\erg\per\second}$, respectively.
For $B=\SI{15}{\micro\gauss}$, as derived by \citet{HESS_LMC_2015} within a leptonic model for \tdorc, the corresponding values are $L_\mathrm{e}^\mathrm{\tdorc}\approx\SI{1.5e37}{\erg\per\second}$ and $L_\mathrm{e}^\mathrm{R136}\approx\SI{9.5e36}{\erg\per\second}$, that is, starting to be dominated by synchrotron losses.

\section{Discussion}\label{sec:discussion}
For most of our discussion, we will assume a connection between the \gam-ray emission and the YMCs LH~90\footnote{While LH~90 is, strictly speaking, classified as an OB association \citep{Testor1993}, we will treat it as a YMC in our discussion, as it fulfills, e.g., the definition of \citet{Vieu2023}.} and R136.
This does not exclude scenarios in which an SN explodes inside the SB formed by the clusters; the inevitable interaction of the SN shock with the YMC and its environment distinguishes this case from a ``normal,'' isolated SNR \citep{Badmaev2024}.
Alternative origins of the \gam-ray emission from the newly detected source \hessjrmc will be discussed at the end of this section.

We provide in Table~\ref{tab:ymsc_properties} an overview of the relevant properties of the YMCs and the surrounding medium, as well as results derived from the \hess \gam-ray observations.
We emphasize that most of the listed properties of the clusters are highly uncertain.
The assumed values should thus be viewed as estimates only.

\begin{deluxetable}{lccc}
  \tablecaption{YMC properties.\label{tab:ymsc_properties}}
  \tablehead{
    \colhead{Property} & \colhead{LH~90} & \colhead{R136} & Ref.
  }
  \startdata
    Cluster age [\si{\mega\year}] & 4 & 1.5 & 1,2\\
    Wind power\tablenotemark{a} [\SI{e38}{\erg\per\second}] & 1.5 & 10 & 3,4\\
    Wind velocity [\si{\kilo\meter\per\second}] & 3000 & 3000 & 5,6\\
    Average ISM density [\si{\per\cubic\centi\meter}] & 100 & 100 & 7--11\\
    Magnetic field [\si{\micro\gauss}] & 15 & 15\tablenotemark{b} & 12\\\hline
    SB radius [pc] & 74 & 56\\
    Termination shock radius [pc] & 7.9 & 8.7\\\hline
    2D Gaussian width [pc] & 27.8 & 33.5\\
    68\% containment radius [pc] & 42.0 & 50.5\\
    Spectral index & $-2.57$ & $-2.54$\\
    Flux\tablenotemark{c} [\SI{e-13}{\per\square\centi\meter\per\second}] & 4.8 & 3.6\\
    Luminosity\tablenotemark{c} [\SI{e35}{\erg\per\second}] & 2.9 & 2.2\\\hline
    Req. power ($pp$) [\SI{e36}{\erg\per\second}] & 1.1 & 2.0\\
    Req. power (IC) [\SI{e37}{\erg\per\second}] & 1.5 & 0.95\\
  \enddata
  \tablecomments{
    Entries in the first section of the table are assumed properties, based on information available in the literature.
    The expected size of the SB and its termination shock -- assuming a spherical expansion -- are derived from these properties, following \citet{Weaver1977} and \citet{Koo1992}.
    Entries in the third section summarize the properties of the \gam-ray emission measured with \hess (cf.\ Sect.~\ref{sec:results}; see Table~\ref{tab:parameters} for fit results including uncertainties).
    The last two rows give the required power in primary CR protons in the hadronic ($pp$) scenario and the required power in primary CR electrons in the leptonic (IC) scenario, respectively, as determined from the \hess measurements.
    For the hadronic case, we used the primary spectra with a break at \SI{1}{TeV} (cf.~Sect.\ \ref{sec:results}), considered no escape or radiation losses, and assumed a continuous injection over the lifetime of the respective cluster.
    \tablenotetext{a}{Averaged over the cluster lifetime.}
    \tablenotetext{b}{No literature estimate available.}
    \tablenotetext{c}{The integrated flux and luminosity are given above an energy of \SI{0.5}{TeV}, which is the energy threshold of the \hess analysis.}
  }
  \tablerefs{(1)~\citealt{Testor1993}; (2)~\citealt{Crowther2016}; (3)~\citealt{Kavanagh2015}; (4)~\citealt{Crowther2010}; (5)~\citealt{Kavanagh2019}; (6)~\citealt{Brands2022}; (7)~\citealt{Sano2017}; (8)~\citealt{Yamane2021}; (9)~\citealt{Johansson1998}; (10)~\citealt{Indebetouw2013}; (11)~\citealt{Kalari2018}; (12)~\citealt{HESS_LMC_2015}.}
\end{deluxetable}

For \tdorc, surrounding LH~90, we note in addition that the presence of a shell of nonthermal X-ray emission implies the existence of a fast-moving ($\gtrsim \SI{3000}{\kilo\meter\per\second}$) shock, which cannot be the forward shock of the expanding SB, as H$\alpha$ observations indicate a shell expanding at only $\lesssim\SI{100}{\kilo\meter\per\second}$.
\citet{Kavanagh2019} have concluded from this that the X-ray emission is due to an SNR shock wave that expands fast in the low-density ($\sim \SI{e-3}{\per\cubic\centi\meter}$) interior of the SB.
An alternative explanation could be the termination shock of the collective wind from the LH~90 association of clusters.

Regarding R136, we are not aware of an estimate for the average magnetic field strength in the surroundings of the cluster.
For ease of comparison, we assume here the same value adopted for \tdorc (\SI{15}{\micro\gauss}; \citealt{HESS_LMC_2015}).
Based on its total mass ($\sim$$\num{2.2e4}\,M_\odot$; \citealt{Cignoni2015}) and compactness, R136 is expected to exhibit a collective wind and inflate an SB \citep[see, e.g.,][]{Vieu2023}, similar to the case of \tdorc.
However, no SB around R136 could yet be unambiguously identified -- although a swept-up shell of gas \citep{Chu1994} as well as diffuse thermal X-ray emission \citep[e.g.][]{Townsley2006} have been detected, which may be attributed to the working of a wind emanating from the cluster.
\citet{Wang1991}, on the other hand, have identified several H$\alpha$ shells that seem to intersect at the position of R136.
The lack of a spherical shell may be attributed to the inhomogeneity of the interstellar medium (ISM) around R136 (cf.\ \citealt{Johansson1998,Kalari2018} and Fig.~\ref{fig:spitzer_ir_image_r136}).
For the purpose of our discussion, we nevertheless compute and list in Table~\ref{tab:ymsc_properties} the expected size of a putative SB around R136 and its termination shock.

It is interesting that in terms of their \gam-ray emission, \tdorc and R136 look similar: both appear extended with a Gaussian width of around \SI{30}{pc} and exhibit a relatively soft spectrum with a spectral index of around $-2.6$.
This is despite the YMCs at their centers being rather unequal: R136 is younger but allegedly exerts a more powerful wind than the LH~90 association.
However, it appears that these differences compensate in the sense that the expected size of the SB and position of the termination shock are comparable between the two cases.
We thus cannot conclude that the \gam-ray emission must originate from different processes or be created at different sites.

In both cases, the \gam-ray emission extends well beyond the expected location of the termination shock of the collective cluster wind.
This may point to a scenario that is different from the case of Westerlund~1, where the \gam-ray emission was found to exhibit a ringlike structure with a radius similar to that of the termination shock \citep{HESS_Wd1_2022}.
We note, however, that compared to Westerlund~1, R136 and \tdorc are located in a region with an ISM density that is, on average, 1 order of magnitude larger.
Therefore, at least in a hadronic scenario, it is still conceivable that CR nuclei accelerated at the wind termination shock and interacting with gas clouds further away from the cluster are responsible for the \gam-ray emission.
In that case, on the other hand, one would expect the centroid of the \gam-ray emission to coincide with the positions of the densest gas clouds (see Figs.~\ref{fig:vlt_optical_image_tdorc} and~\ref{fig:spitzer_ir_image_r136}), whereas we find it to lie very close to the position of the YMC for both \tdorc and R136, somewhat disfavoring a hadronic origin of the emission.

The feasibility of the leptonic and hadronic emission scenario can also be scrutinized by comparing the power in primary CRs required to sustain the \gam-ray emission (see Table~\ref{tab:ymsc_properties}) with the power provided by, for example, the cluster wind, noting that the large uncertainties associated with either prevent a detailed discussion.
For \tdorc, we obtain in the leptonic scenario a ratio between these two quantities of $\sim$10\%.
While this would be a surprisingly large efficiency for a leptonic accelerator, we stress that the wind power listed in Table~\ref{tab:ymsc_properties} (\SI{1.5e38}{\erg\per\second}) is an estimate for the \emph{average} power over the cluster lifetime; \citet{Kavanagh2015} have estimated that the \emph{current} power of just the known Wolf--Rayet stars in LH~90 is $\sim\SI{5e38}{\erg\per\second}$, which alleviates the requirements slightly.
Nevertheless, it appears challenging in this scenario to entirely explain the \gam-ray emission as resulting from the collective wind of the clusters.
As already proposed previously, a recent SN in the LH~90 association may be providing the additional energy that is required to explain the \gam-ray signal \citep{HESS_LMC_2015,Kavanagh2019}.
For R136, with its more powerful wind, we find a less demanding but still considerable efficiency of $\sim$1\% in the leptonic case.

Considering, on the other hand, a hadronic origin of the \gam-ray emission, we deduce a minimum ratio between required power in protons and power provided by the cluster wind of $\sim$0.7\% for \tdorc and of $\sim$0.2\% for R136.
We assume in this case that the \gam-ray emission originates from interactions of CRs in the dense gas clouds that surround the respective SBs (see Figs.~\ref{fig:vlt_optical_image_tdorc} and~\ref{fig:spitzer_ir_image_r136}); this is in line with the extension of the \gam-ray emission approximately matching the expected size of the SB.
The derived acceleration efficiencies are considerably lower than those typically derived in the framework of diffusive shock acceleration (DSA), which are $\mathcal{O}$(10\%) \citep[e.g.][]{Eichler1979}.
In terms of energy requirements, the hadronic scenario thus appears viable, even for somewhat lower gas densities than assumed here.
However, a hadronic scenario for \tdorc is disfavored, as it requires relatively large magnetic field strengths, which is in disagreement with the magnetic field estimate by \citet{Kavanagh2019}.
We also note that for both sources, a mixed leptonic--hadronic scenario is possible.

For the leptonic models, the inferred electron distributions ($\Gamma_\mathrm{e}\approx -3.2$) are consistent with a synchrotron-cooled injection spectrum of $\Gamma_\mathrm{e,inj}\approx -2.2$, close to the standard prediction of DSA.
In the hadronic scenario, the derived spectral indices for the proton distributions ($\Gamma_\mathrm{p}\approx -2.6$), while in line with those inferred from other massive star clusters, are steeper than the DSA prediction.
This could be explained by energy-dependent escape from the emitting region, though it has been argued that steep spectra can also result from the acceleration process at wind termination shocks; see, for example, \citet{Webb1985}.
Future \gam-ray observations may disentangle these different possibilities \citep[see, e.g.,][]{CTA2023}.

Regarding the multiwavelength picture, one striking difference between \tdorc and R136 is the presence of a nonthermal X-ray shell in the former and the lack thereof in the latter (although a diffuse X-ray source coincident with R136 has been detected with eROSITA; \citealt{Sasaki2022}).
This may be seen as a hint for a different origin of the \gam-ray emission: R136 is too young for many stars to have exploded yet, and so its emission may be from stellar winds alone, whereas \tdorc may be powered by a combination of winds and a recent SN.
There are, however, older massive stars in the encompassing cluster NGC~2070 \citep{Sabbi2012,Bestenlehner2020}, still within the putative SB of R136, and so SNe should have occurred within this region during the past $\sim\SI{e6}{\year}$, which calls for a different explanation for the dissimilar appearance of \tdorc and R136 in the X-ray domain.

Finally, we comment on the possibility of the \gam-ray emission from \hessjrmc not being connected to R136 (i.e.\ neither to the collective cluster wind nor to SNe occurring inside the SB around the cluster).
As PWNe constitute a large fraction of extended TeV \gam-ray sources, it is natural to consider an association of \hessjrmc with a PWN.
While this generally appears realistic in terms of the source extension and energy spectrum, there is no evidence for the presence of a PWN that could plausibly be associated with R136.
It seems unlikely that a pulsar/PWN with a power output of $\sim \SI{e37}{\erg\per\second}$ (cf.\ Sect.~\ref{sec:results}) leaves no trace at any other wavelength, despite the Tarantula Nebula being one of the most deeply observed regions in any wave band.
Another theoretically viable explanation would be a counterpart to \hessjrmc that is located in front of or behind R136 along the line of sight.
However, as there is again no evidence of such an object, the association with R136 appears more likely.

\section{Conclusion}
We present new measurements of the VHE \gam-ray emission from the Tarantula Nebula region in the LMC with the \hess array of Cherenkov telescopes.
Utilizing improved analysis techniques, we are able to resolve the emission into three distinct \gam-ray sources.
The brightest one, \hessjpwn, is associated with the PWN N~157B.
The other two sources can be associated with YMCs and/or their surrounding SBs.
Both sources are extremely luminous in \gam-rays, exceeding even the most massive Galactic young star cluster Westerlund~1 in this regard.

We provide updated results on \hessjsb, associated with the SB \tdorc around the LH~90 OB association.
The extension of the \gam-ray emission -- measured for the first time -- is comparable to the size of the nonthermal X-ray shell around LH~90, suggestive of a common origin.
A consideration of the energy requirements suggests that the emission is powered by a recent SN in this case.
A lack of correlation between the \gam-ray emission and the distribution of molecular gas disfavors a hadronic origin, although we cannot rule out hadronic contributions.

Furthermore, we report the discovery of a new \gam-ray source associated with the YMC R136, labeled \hessjrmc.
This source is similar in terms of spatial extension and \gam-ray energy spectrum to the source associated with \tdorc.
However, the lack of an identified SB around R136 complicates the interpretation in this case.
Given that R136 is likely to exhibit a strong collective cluster wind, both a leptonic and a hadronic origin of the \gam-ray emission appear viable.

The detection of \gam-ray emission from the direction of R136 adds to the growing list of YMCs associated with TeV emission.
Despite still being small, the population shows quite some variety, in terms of both the \gam-ray emission and its interpretation.
Our analysis thus provides crucial information to understand the ability of YMCs to accelerate CRs better.

\begin{acknowledgments}
\section*{Acknowledgements}
{\small
We thank Hidetoshi Sano and Yumiko Yamane for providing gas maps of the \tdorc region.

The support of the Namibian authorities and of the University of Namibia in facilitating the construction and operation of \hess is gratefully acknowledged, as is the support by
the German Ministry for Education and Research (BMBF),
the Max Planck Society,
the Helmholtz Association,
the French Ministry of Higher Education, Research and Innovation,
the Centre National de la Recherche Scientifique (CNRS/IN2P3 and CNRS/INSU),
the Commissariat \`{a} l'\'{e}nergie atomique et aux \'{e}nergies alternatives (CEA),
the U.K. Science and Technology Facilities Council (STFC),
the Irish Research Council (IRC) and the Science Foundation Ireland (SFI),
the Polish Ministry of Education and Science, agreement No.~2021/WK/06,
the South African Department of Science and Innovation and National Research Foundation,
the University of Namibia,
the National Commission on Research, Science \& Technology of Namibia (NCRST),
the Austrian Federal Ministry of Education, Science and Research and the Austrian Science Fund (FWF),
the Australian Research Council (ARC),
the Japan Society for the Promotion of Science,
the University of Amsterdam, and
the Science Committee of Armenia grant 21AG-1C085.
We appreciate the excellent work of the technical support staff in Berlin, Zeuthen, Heidelberg, Palaiseau, Paris, Saclay, T\"{u}bingen, and Namibia in the construction and operation of the equipment.
This work benefited from services provided by the \hess Virtual Organisation, supported by the national resource providers of the EGI Federation.
\software{
  Gammapy \citep{Donath2023,Deil2020},
  Astropy \citep{Robitaille2013,PriceWhelan2018,PriceWhelan2022},
  matplotlib \citep{Hunter2007},
  iminuit \citep{Dembinski2020}.
}
}
\end{acknowledgments}

\clearpage

\appendix

\section{Analysis Procedure and Fit of Hadronic Background Model}
\label{sec:appx_bkg_fit}
In this appendix, we provide technical details about the \hess data analysis. 
Furthermore, we present the results of fitting the model for the residual hadronic background to the observation runs.

The likelihood analysis is carried out in a three-dimensional geometry with specifications as listed in Table~\ref{tab:analysis_settings}.
For every bin $i$ in this ``cube,'' we calculate a number of expected events $\mu_i$, where we take into account contributions from the hadronic background model as well as from \gam-ray source models, if present.
The prediction for the latter is obtained by folding the spatial and spectral source model with the IRFs, that is, with the exposure, energy dispersion matrix, and point-spread function (PSF).
The fit then proceeds by comparing $\mu_i$ to the number of actually observed events, $n_i$, simultaneously across all bins.
More specifically, the best-fit models are obtained by adjusting the model parameters such that the quantity $-2\log(\mathcal{L})$ is minimized, with the likelihood $\mathcal{L}=\prod_i P(n_i|\mu_i)$ and $P(n_i|\mu_i)$ denoting the Poisson probability to observe $n_i$ events, given an expectation of $\mu_i$.
The minimization is done numerically, where we have used the default fitting backend in \gammapy, \textsc{iminuit} \citep{Dembinski2020}.
Two different models with optimized likelihoods $\mathcal{L}_0$ and $\mathcal{L}_1$ can be compared by means of a likelihood ratio test.
In the limit of sufficient statistics and in case the parameter values are far from boundaries, the test statistic $\mathrm{TS}=-2\log(\mathcal{L}_0/\mathcal{L}_1)$ follows a $\chi^2$ distribution with $k$ degrees of freedom if the two models are nested (i.e.\ model~1 can always be reduced to model~0 for a particular choice of parameter values), where $k$ is the difference in the number of model parameters between the models \citep{Wilks1938}.

\begin{deluxetable}{cc}
  \tablecaption{Summary of analysis settings.\label{tab:analysis_settings}}
  \tablehead{\colhead{Setting} & \colhead{Value}}
  \startdata
    \multirow{2}{*}{ROI center (J2000)} & R.A.~\hms{5}{35}{28.25}\\
     & Decl.~$-69^\circ 16'13.08''$\\
    ROI size & $5^\circ\times 5^\circ$\\
    Spatial pixel size & $0.02^\circ\times 0.02^\circ$\\
    Maximum offset angle & $1.5^\circ$\\
    Energy binning & 16 bins decade$^{-1}$\\
    Energy range & \SI{0.5}{TeV}--\SI{100}{TeV}\\
  \enddata
  \tablecomments{The maximum offset angle denotes the maximum allowed angle between the reconstructed direction of every event and the pointing direction of the telescopes.}
\end{deluxetable}

\begin{figure*}
  \centering
  \subfigure[]{
    \includegraphics[width=0.48\textwidth]{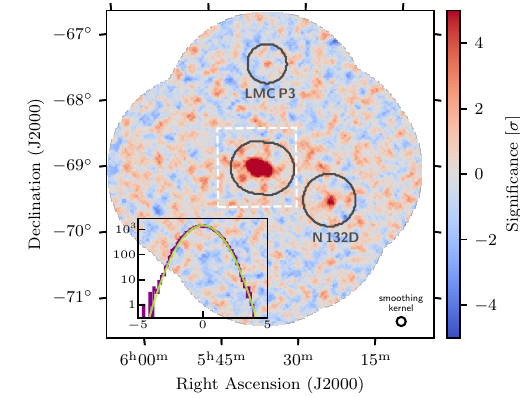}
    \label{fig:sign_map_bkg_fit}
  }
  \subfigure[]{
    \includegraphics[width=0.48\textwidth]{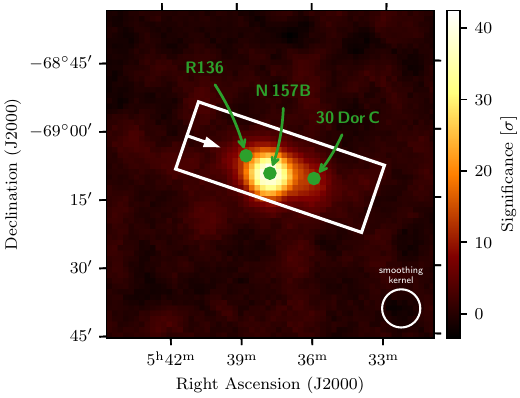}
    \label{fig:sign_map_slice_box}
  }
  \caption{
    (a) Significance map for the entire ROI, above an energy of \SI{0.5}{TeV}, after adjustment of the hadronic background model.
    The map has been smoothed with a top-hat kernel of $0.07^\circ$ radius, as indicated in the bottom right corner.
    The dark gray lines enclose regions that have been excluded in the fit of the background, the white dashed line indicates the target region of the analysis, shown in more detail on the right.
    The inset shows the distribution of significance values outside the exclusion regions (purple histogram) as compared to a normal distribution with unity width (green line).
    (b) Cutout of the significance map, showing the target region.
    The white rectangle denotes a slice along which \gam-ray excess profiles have been computed (cf.\ Fig.~\ref{fig:excess_profiles}).
  }
  \label{fig:sign_maps}
\end{figure*}

Although care is taken in the construction of the hadronic background model to predict the background rate in every given observation run as accurately as possible, it is typically necessary to adjust the model slightly to the actually observed level of background in each run \citep[see][]{Mohrmann2019}.
To avoid a bias due to actual \gam-ray emission in this procedure, we mask regions in the ROI that contain known \gam-ray sources in this step (i.e.\ the corresponding spatial pixels do not enter the likelihood computation).
Besides the sources discussed in detail in this work, this includes the SNR N~132D \citep{HESS_N132D_2021} and the \gam-ray binary LMC~P3 \citep{HESS_LMCP3_2018}.
For every observation run, we fit two parameters of the background model: a global normalization parameter $\phi$ that scales the total background rate and a ``spectral tilt'' parameter $\delta$ that modifies the predicted rate $R$ at energy $E$ as $R'=R\cdot (E/\SI{1}{TeV})^{-\delta}$.
We show in Fig.~\ref{fig:sign_map_bkg_fit} an energy-integrated residual significance map for the full ROI obtained after the background model adjustment, where we have summed the predicted background and observed events of all observations and use the ``Cash'' statistic \citep{Cash1979} to determine the residual significance in each spatial pixel.
In the absence of systematic biases, the distribution of significance values in all pixels outside the masked regions should follow a Gaussian distribution centered at zero and with a width of unity, reflecting statistical fluctuations around the expected rate.
The inset in Fig.~\ref{fig:sign_map_bkg_fit} shows that this is very nearly the case for this analysis, indicating that an excellent description of the hadronic background has been achieved.
In Fig.~\ref{fig:sign_map_slice_box}, we provide a zoom-in of the target region of this work, with the location of three detected \gam-ray sources indicated (cf.\ Sect.~\ref{sec:results} and Appendix~\ref{sec:appx_src_fit}).

For the subsequent modeling of the \gam-ray emission in the target region, we divide (for technical reasons) the observations into six groups, where the first group contains all observations taken between 2004 and 2013 and the remaining observations are grouped into yearly observation periods.
For each group, a ``stacked'' data set is created by summing the observed number of events, exposure, and predicted background events and averaging the energy dispersion matrix and PSF.
The source modeling is then performed as a joint likelihood analysis across these six stacked data sets.

\newpage
\section{Source Fitting}
\label{sec:appx_src_fit}
In this appendix, we detail the modeling of the \gam-ray sources in the Tarantula Nebula region as displayed in Fig.~\ref{fig:flux_maps}.
All best-fit parameter values of the final ROI model are summarized in Table~\ref{tab:parameters}.
In what follows, we first describe the modeling procedure and its main results before we introduce our method of estimating systematic uncertainties for the fit parameters.

\begin{deluxetable*}{ccc}
  \tablecaption{Best-fit parameters of the \gam-ray source models.\label{tab:parameters}}
  \tablehead{\colhead{Parameter} & \colhead{Unit} & \colhead{Value}}
  \startdata
    \multicolumn3c{\textbf{N\,157B / \hessjpwn}}\\\hline
    R.A. & deg & $84.4394\pm 0.0048\estat$ (\hmse{5}{37}{45.5}{1.1})\\
    Decl. & deg & $-69.1713\pm 0.0016\estat$ ($-69^\circ 10'17''\pm 6_\mathrm{stat}''$)\\
    $\sigmaext$ & deg & $0.0137\pm 0.0033\estat \pm 0.0030\esys$\\
    $\phi_0$ & $10^{-13}\,\mathrm{TeV}^{-1}\,\mathrm{cm}^{-2}\,\mathrm{s}^{-1}$ & $8.69\pm 0.56\estat \pm 0.85\esys$\\
    $\alpha$ & --- & $2.03\pm 0.07\estat \pm 0.08\esys$\\
    $\beta$ & --- & $0.311\pm 0.037\estat$\\
    \hline
    \multicolumn3c{\textbf{\tdorc\ / \hessjsb}}\\\hline
    R.A. & deg & $84.021\pm 0.018\estat$ (\hmse{5}{36}{5.0}{4.3})\\
    Decl. & deg & $-69.197\pm 0.006\estat$ ($-69^\circ 11'49''\pm 22_\mathrm{stat}''$)\\
    $\sigmaext$ & deg & $0.0319\pm 0.0066\estat \pm 0.0034\esys$\\
    $\phi_0$ & $10^{-13}\,\mathrm{TeV}^{-1}\,\mathrm{cm}^{-2}\,\mathrm{s}^{-1}$ & $2.54\pm 0.37\estat\,_{-0.40}^{+0.44}|\esys$\\
    $\Gamma$ & --- & $2.57\pm 0.09\estat$\\
    \hline
    \multicolumn3c{\textbf{R136 / \hessjrmc}}\\\hline
    R.A. & deg & $84.692\pm 0.038\estat$ (\hmse{5}{38}{46}{9})\\
    Decl. & deg & $-69.103\pm 0.013\estat$ ($-69^\circ 06'11''\pm 47_\mathrm{stat}''$)\\
    $\sigmaext$ & deg & $0.0384\pm 0.0090\estat \,_{-0.0037}^{+0.0045}|\esys$\\
    $\phi_0$ & $10^{-13}\,\mathrm{TeV}^{-1}\,\mathrm{cm}^{-2}\,\mathrm{s}^{-1}$ & $1.90\pm 0.58\estat\,_{-0.38}^{+0.45}|\esys$\\
    $\Gamma$ & --- & $2.54\pm 0.15\estat$\\
  \enddata
  \tablecomments{
    Coordinates are given for the epoch J2000.
    The flux normalization $\phi_0$ is specified at a reference energy of $E_0=\SI{1}{TeV}$.
    None of the sources exhibit significant emission above \SI{30}{TeV} (cf.\ Figs.~\ref{fig:luminosities} and~\ref{fig:seds}).
    Statistical errors are at a 68\% confidence level.
    Systematic errors have been derived as explained in Sect.~\ref{sec:sys_err}; no systematic errors are quoted in case they are found to be negligible with respect to the statistical error.
    Systematic errors do not include a systematic uncertainty in the pointing of the telescopes, which can be of the order of $10''$--$20''$ \citep{Gillessen2004}.
  }
\end{deluxetable*}

\subsection{Modeling Procedure and Results}\label{sec:model_results}
\begin{figure*}
  \centering
  \includegraphics{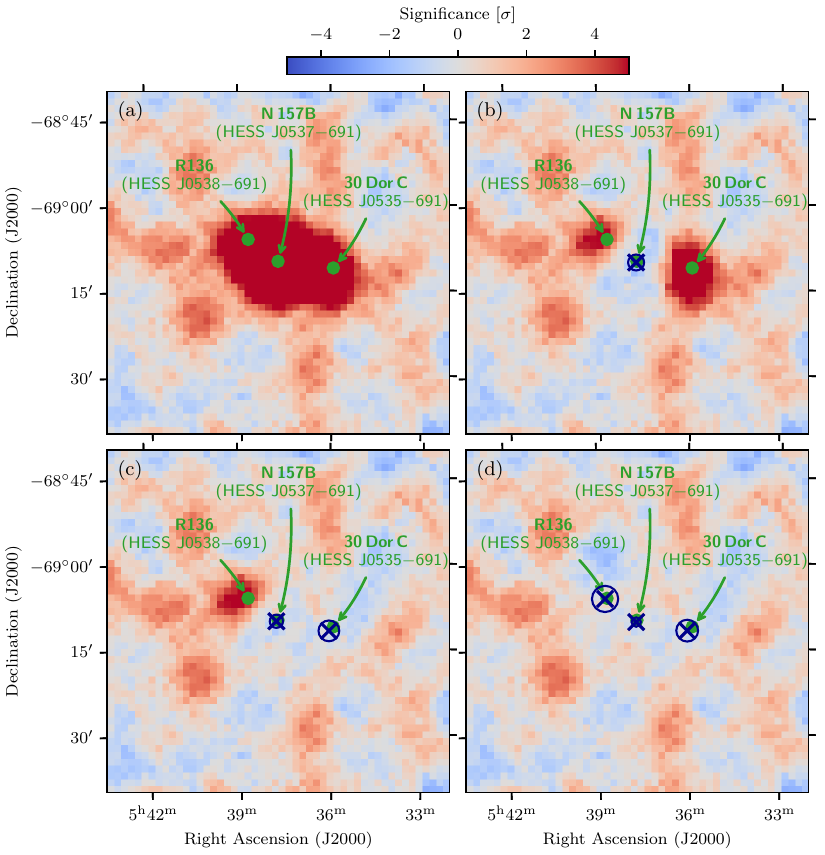}
  \caption{
    Maps of residual significance of the analysis target region.
    (a) After the background model adjustment (cf.\ Appendix~\ref{sec:appx_bkg_fit}).
    (b) After adding N\,157B to the ROI model.
    (c) After adding \tdorc to the ROI model.
    (d) After adding R136 to the ROI model.
    The blue crosses denote the best-fit position of the respective source models, while the blue circles indicate the best-fit Gaussian width.
  }
  \label{fig:sign_maps_models}
\end{figure*}

The iterative modeling procedure is illustrated in Fig.~\ref{fig:sign_maps_models}.
For all components, we use as a spatial model a two-dimensional, symmetric Gaussian with width parameter $\sigmaext$.
The positions of the models are free to vary, i.e.\ not fixed to the nominal positions of the respective counterparts.

\begin{figure*}
  \centering
  \subfigure[]{
    \includegraphics[width=0.48\textwidth]{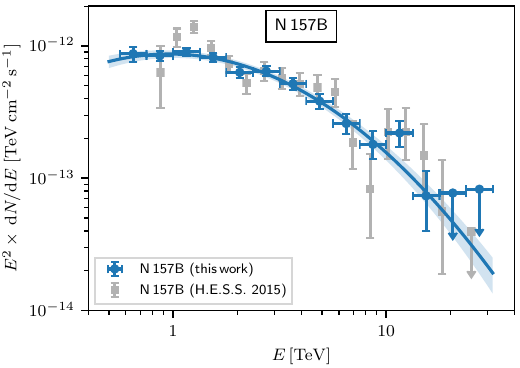}
    \label{fig:sed_n157b}
  }
  \subfigure[]{
    \includegraphics[width=0.48\textwidth]{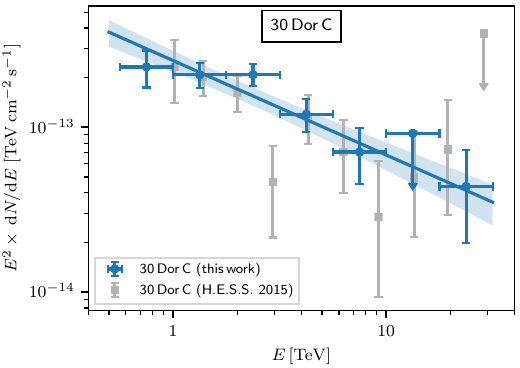}
    \label{fig:sed_tdorc}
  }
  \subfigure[]{
    \includegraphics[width=0.48\textwidth]{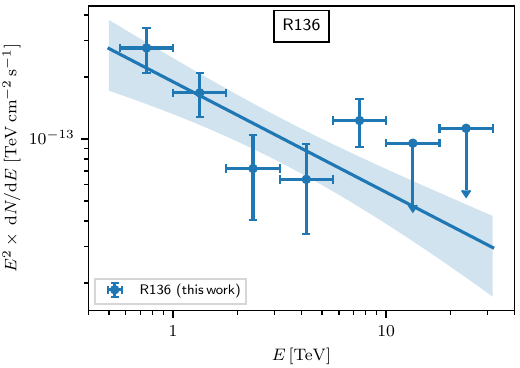}
    \label{fig:sed_r136}
  }
  \caption{
    Spectral energy distribution for N\,157B (a), \tdorc (b), and R136 (c).
    The blue line and band show the best-fit spectral model with its statistical uncertainty.
    The flux points have been obtained by refitting the flux normalization $\phi_0$ in the corresponding energy range, keeping the other parameters of the spectral model of the respective source fixed.
    The published spectra are taken from \citet{HESS_LMC_2015}.
  }
  \label{fig:seds}
\end{figure*}

\begin{figure}
  \centering
  \includegraphics{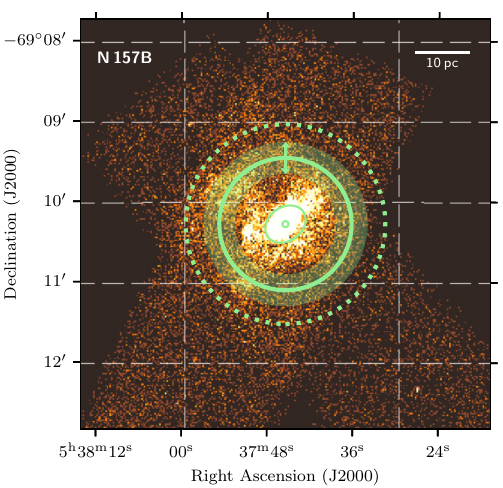}
  \caption{
    Chandra X-ray image of N\,157B \citep{Chen2006}, with properties of the best-fit \gam-ray source model overlaid.
    The green circle marker, thin solid line, thick solid line, and thick dotted line indicate the best-fit position, position uncertainty (95\% confidence level), 1-$\sigma$ Gaussian radius, and 68\% containment radius, respectively.
    In addition, the transparent band shows the statistical uncertainty on the 1-$\sigma$ radius, while the arrows denote the systematic uncertainty.
  }
  \label{fig:chandra_xray_image_n157b}
\end{figure}

The first source we include in our model is the PWN N\,157B \citep{HESS_N157B_2012,HESS_LMC_2015}.
Its energy spectrum, shown in Fig.~\ref{fig:sed_n157b}, is modeled with a log-parabola function,
\begin{equation}\label{eq:logparabola}
  \frac{\mathrm{d}N}{\mathrm{d}E} = \phi_0\left(\frac{E}{E_0}\right)^{-\alpha-\beta\ln(E/E_0)}\,,
\end{equation}
where the reference energy $E_0=\SI{1}{TeV}$ remains fixed in the fit.
The log-parabola model is preferred over a simple power-law model with a significance of 7.5$\sigma$.
As shown in Fig.~\ref{fig:chandra_xray_image_n157b}, the best-fit position of the model is in excellent agreement with the center of the PWN as seen in X-rays.

After the inclusion of N\,157B, the largest excess emission is visible around \tdorc\ (cf.\ Fig.~\ref{fig:sign_maps_models}(b)), which we add as a second source to our model.
This leads to an improvement in the fit statistic of $\mathrm{TS}\approx 138.9$, which -- assuming 5 additional degrees of freedom for the model -- translates to a detection significance of 11$\sigma$.
The source appears significantly extended; a pointlike source is excluded with 3.3$\sigma$ significance.
We model the energy spectrum with a power-law function,
\begin{equation}\label{eq:powerlaw}
  \frac{\mathrm{d}N}{\mathrm{d}E} = \phi_0\left(\frac{E}{E_0}\right)^{-\Gamma}
\end{equation}
(with $E_0=\SI{1}{TeV}$ again fixed), and find it to be in good agreement with our earlier measurement \citep{HESS_LMC_2015}, as shown in Fig.~\ref{fig:sed_tdorc}.
A log-parabola model for \tdorc improves the fit only marginally ($\mathrm{TS}\approx 4.2$) and is thus not selected as the default model.

Figure~\ref{fig:sign_maps_models}(c) shows that residual emission is still visible after the addition of \tdorc, close to the location of the star cluster R136.
Adding a model component for this source improves the fit by $\mathrm{TS}\approx 53.6$, implying a detection significance of 6.3$\sigma$.
The extended source spatial model is preferred with respect to a pointlike model by 3.1$\sigma$.
As for \tdorc, the spectral model is a power-law function, displayed in Fig.~\ref{fig:sed_r136}.

\begin{figure*}
  \centering
  \includegraphics{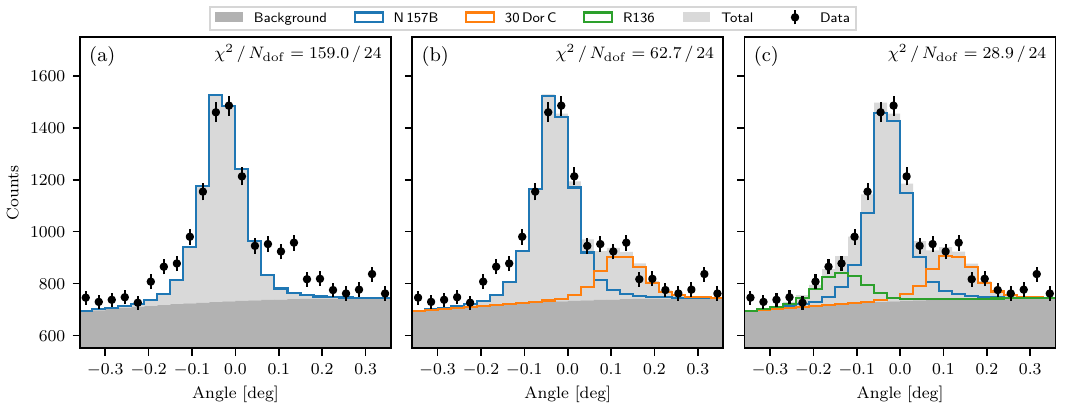}
  \caption{
    Count profiles for the slice across the target region indicated in Fig.~\ref{fig:sign_map_slice_box}.
    The black data points, identical in all panels, show the observed number of events in $0.03^\circ$ wide bins along the slice.
    The dark gray histogram displays the predicted residual hadronic background, while the light gray histogram denotes the total model prediction including \gam-ray source models.
    (a) After adding N\,157B to the ROI model.
    (b) After adding \tdorc to the ROI model.
    (c) After adding R136 to the ROI model.
    Indicated in the top right corner of each panel is the result of a $\chi^2$ test between the total model prediction and the observed data.
  }
  \label{fig:excess_profiles}
\end{figure*}

With N\,157B, \tdorc, and R136 included in the ROI model, the residual significance map (see Fig.~\ref{fig:sign_maps_models}(d)) shows no further significant excess emission.
The iterative modeling procedure hence stops at this point.
As a demonstration of the good agreement between the observed data and the final model, we show in Fig.~\ref{fig:excess_profiles} one-dimensional profiles along the slice displayed in Fig.~\ref{fig:sign_map_slice_box}.
For our final model, a comparison between the model prediction and the observed data by means of a $\chi^2$ test yields $\chi^2=28.9$ for 24~degrees of freedom.

We have modified our ROI model in various ways to test whether a significantly better-fitting model can be found.
For example, as already alluded to in the main text, we have replaced the Gaussian spatial model for N\,157B with a pointlike one.
Surprisingly, given the relatively small statistical error of $0.0033^\circ$ (at a 68\% confidence level) on the measured extension of $\sigmaext^\mathrm{N\,157B}=0.0137^\circ$, we find the pointlike model to be disfavored by only $\approx$1.3$\sigma$.
We attribute this to the presence of the two other \gam-ray sources, \tdorc and R136, which could be more extended than our best-fit model suggests, thus causing the likelihood profile for the extension parameter of N\,157B to flatten toward smaller values.
Despite the marginal preference for the Gaussian spatial model for N\,157B, we use it in our final model in order to not obtain biased results for the extension of \tdorc or R136.
For reference, if N\,157B is modeled as a pointlike source, we obtain $\sigmaext^\mathrm{\tdorc}=(0.0334\pm 0.0066\estat)^\circ$ and $\sigmaext^\mathrm{R136}=(0.0431\pm 0.0089\estat)^\circ$, which is compatible with our default result (cf.\ Table~\ref{tab:parameters}) within the uncertainties.
Interpreted as an upper limit, our analysis yields $\sigmaext^\mathrm{N\,157B}<0.0190^\circ$ at a 95\% confidence level (statistical uncertainties only).

We have also tested a model in which the Gaussian spatial model for N\,157B is allowed to be elongated.
While this improves the fit quality considerably when only N\,157B is included as a \gam-ray source ($\mathrm{TS}\approx 9.1$), this is no longer the case when \tdorc and R136 are included as well ($\mathrm{TS}\approx 0.5$).
Conversely, the model with all three sources included is strongly preferred over that with only N\,157B, even when modeled as elongated ($\mathrm{TS}\approx 183.4$; note, however, that the two models are not nested in this case).
This is still true when adding a component for \tdorc\ (in this case $\mathrm{TS}\approx 50$), implying that also just the emission around R136 cannot be explained by allowing the model for N\,157B to be elongated.

\subsection{Estimation of Systematic Uncertainties}
\label{sec:sys_err}
In this section, we explain our procedure for estimating the systematic uncertainties on the \gam-ray source model parameters specified in Table~\ref{tab:parameters}.
We study three different sources of systematic errors: a mismodeling of the instrument PSF, an incorrect estimation of the residual hadronic background, and a wrongly calibrated energy scale.
For each effect, we determine the associated systematic uncertainty following a ``bracketing'' approach.
That is, we vary the IRFs according to the assumed magnitude of the effect, repeat the source modeling, and quote the difference to the best-fit parameter values obtained with the nominal IRFs as systematic error.
The total uncertainties stated in Table~\ref{tab:parameters} have been computed by summing quadratically the errors obtained for each of the three effects.
Where no systematic uncertainty is specified, the error derived with our method is negligible compared to the statistical uncertainty.
This does not necessarily mean that there is no systematic uncertainty on the respective parameter in general.
For instance, the fitted source positions are subject to possible pointing inaccuracies of the telescopes, which can be of the order of $10''$--$20''$ \citep{Gillessen2004} and are not covered by the systematic effects we studied in detail here.

\subsubsection{PSF}
We validate the accuracy of the PSF for the analysis configuration employed in this study using the bright \gam-ray source \pks, an active galactic nucleus that appears pointlike for \hess\ \citep[e.g.][]{HESS_PKS2155_2005,HESS_PKS2155_2010}.
Using the same configuration, we analyze 576~observation runs (worth $\approx$\SI{274.2}{h} of observation time) on \pks taken between 2004 July 14 and 2012 November 10 above an energy threshold of \SI{0.27}{TeV}.
Observations taken during the very strong outburst of \pks in July 2006 \citep{HESS_PKSFlare_2007} have been excluded from the list of analyzed runs, such that the outcome is not dominated by observations taken over a few days only.
We apply the same procedure of adjusting the hadronic background model to each observation run, as explained in Appendix~\ref{sec:appx_bkg_fit}.
Subsequently, all observations are combined to obtain a stacked data set.

\begin{figure*}
  \centering
  \includegraphics{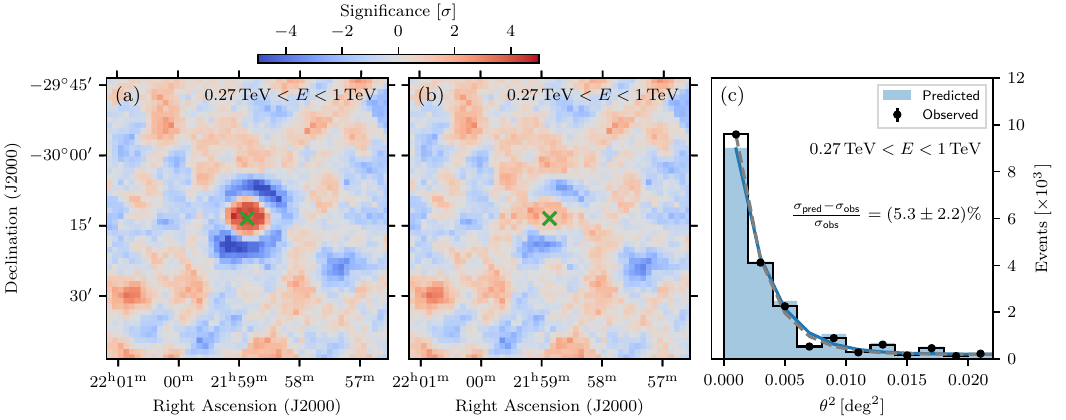}
  \caption{
    Evaluation of systematic uncertainties of the PSF using \pks data.
    (a) Residual significance map obtained with the nominal PSF, in the energy range \SIrange{0.27}{1}{TeV}.
    (b) Same but with the PSF made narrower by 5\%.
    In both maps, the green cross marks the position of \pks, which is modeled as a pointlike source.
    (c) Distribution of $\theta^2$, where $\theta$ is the angular offset between the reconstructed direction of each event and the position of \pks.
    The black histogram and data points show the observed data, while the blue histogram displays the expected distribution based on the nominal PSF.
    The dashed gray and solid blue lines show fits of a Gaussian to the two distributions, respectively.
    These fits indicate that the predicted distribution is about 5\% broader than the observed one (in terms of $\theta$, not $\theta^2$).
  }
  \label{fig:psf_pks2155}
\end{figure*}

We first model \pks using a pointlike spatial source model, using the nominal PSF of the data set.
As a spectral model, we use a log-parabola function (cf.\ Eq.~\ref{eq:logparabola}).
Inspecting the residual significance map for energies between \SI{0.27}{TeV} and \SI{1}{TeV}, shown in Fig.~\ref{fig:psf_pks2155}(a), we find that the model based on the nominal PSF is not able to describe the data perfectly, indicating a systematic mismodeling of the PSF.
Relative to the total emission, the remaining residuals are small: in a map computed without including a source model for \pks, the peak significance exceeds 250$\sigma$.
The steeply falling \gam-ray spectrum of \pks (we obtain $\alpha=3.63\pm 0.02$) leads to a much smaller excess of \gam-ray events at higher energies, meaning that the systematic effect -- if still present there -- is no longer visible.

To quantify the effect, we fit a Gaussian to the observed and predicted distribution of angular offsets with respect to the position of \pks (see Fig.~\ref{fig:psf_pks2155}(c)), finding that the predicted distribution, based on the nominal PSF, is about 5\% too broad.
As a verification, we repeat the analysis with a PSF that has been artificially narrowed (by scaling the radial axis) by 5\%.
Indeed, the residual significance map obtained from this, displayed in Fig.~\ref{fig:psf_pks2155}(b), is compatible with statistical fluctuations only.
We conclude that a mismodeling of the PSF at a level of 5\% with respect to its width should be taken into account as a systematic uncertainty.

We therefore repeat the analysis of the Tarantula Nebula region, scaling the PSF for each data set such that it becomes narrower/broader by 5\%.
Affected most by this effect are the measured extensions ($\sigmaext$) of the modeled \gam-ray sources, for which we find PSF-related systematic uncertainties of $\pm 0.0025^\circ$ (N\,157B), $\pm 0.0033^\circ$ (\tdorc), and $(_{-0.0035}^{+0.0044}\,)^\circ$ (R136).
A comparison with the total systematic errors (see Table~\ref{tab:parameters}) shows that the latter are strongly dominated by the PSF uncertainty.

\subsubsection{Residual Hadronic Background}
Refitting the normalization of the residual hadronic background model for each of the stacked data sets shows that variations of the order of 0.5\% are still possible.
We therefore repeat the source modeling with the background model normalization varied by $\pm0.5\%$ for all data sets simultaneously.
Variations larger than this are possible for each individual observation run.
These, however, are expected to average out when combining the observations into stacked data sets.
The systematic errors resulting from this estimation are always smaller than the statistical errors.
The largest effect is observed for the spectrum normalization parameters ($\phi_0$), where we find uncertainties of 0.6\% (N\,157B), 3.5\% (\tdorc), and 4.2\% (R136).

\subsubsection{Energy Scale}
Lastly, we study the effect of a wrongly calibrated energy scale.
To this end, we evaluate the effective area and energy dispersion matrix at energies that are scaled up or down by 10\%.
Such a miscalibration of the energy scale can arise, for example, from variations of the aerosol level in the atmosphere \citep{Hahn2014}.
Similarly to the variation of the residual hadronic background, the energy scale mostly affects the spectrum normalization parameters.
We find uncertainties of 10\% (N\,157B), 17\% (\tdorc), and 18\% (R136).
The energy-scale-related uncertainties are thus of the same magnitude as the statistical errors.

\vspace{3.0cm}
\section{Fit of Primary Particle Spectra}
\label{sec:appx_naima_results}
To fit primary particle spectra, we employ the \textsc{naima} package \citep{Zabalza2015}.
Specifically, we use the \texttt{NaimaSpectralModel} wrapper class in \gammapy, which allows us to fit the models to our binned data sets directly (as opposed to fitting them to flux points derived from these).
For both \tdorc and R136, we fit a leptonic model in which the \gam-ray emission is due to IC emission from CR electrons and a hadronic model in which the \gam-rays are produced in interactions of CR nuclei with gas.
As the energy spectrum for the primary particles, we use a power-law model (see Eq.~\ref{eq:powerlaw}), with $E_0=\SI{3}{TeV}$ as a fixed parameter.
For the leptonic models, we include target photon fields as described in Appendix~\ref{sec:appx_ic_targets}.
For the hadronic models, we use in the fit an arbitrarily chosen gas density of \SI{1}{\per\cubic\centi\meter}.
The spectrum normalization parameter ($\phi_0$) is inversely proportional to the gas density, such that the results can easily be scaled to different densities.
The hadronic model is furthermore based on the parameterization of \gam-ray production cross sections presented in \citet{Kafexhiu2014}, which includes a nuclear enhancement factor that is around 1.7 at TeV energies.
In view of the LMC exhibiting a lower metallicity than the Milky Way \citep[e.g.][]{Choudhury2016}, this factor might be slightly too large for the environments studied in this work, which would imply slightly larger spectrum normalization parameters.
The fit is again carried out in three dimensions, with the same spatial models as used in our previous fit (cf.\ Sect.~\ref{sec:model_results}).
The parameters of the spatial models are essentially identical to those obtained previously and thus not reported again here.
The primary particle spectrum model parameters are summarized in Table~\ref{tab:naima_parameters}.

\begin{deluxetable}{ccc}[bh]
  \tablecaption{Best-fit parameters of the primary particle spectrum models.\label{tab:naima_parameters}}
  \tablehead{\colhead{Parameter} & \colhead{Unit} & \colhead{Value}}
  \startdata
    \multicolumn3c{\textbf{\tdorc\ (Leptonic)}}\\\hline
    $\phi_0$ & $10^{34}\,\mathrm{eV}^{-1}$ & $10.1\pm 1.8\estat$\\  % 9.5
    $\Gamma$ & --- & $3.27\pm 0.11\estat$\\  % 3.30
    \hline
    \multicolumn3c{\textbf{R136 (Leptonic)}}\\\hline
    $\phi_0$ & $10^{34}\,\mathrm{eV}^{-1}$ & $7.0\pm 2.4\estat$\\   % 6.7
    $\Gamma$ & --- & $3.19\pm 0.17\estat$\\  % 3.22
    \hline
    \multicolumn3c{\textbf{\tdorc\ (Hadronic)}}\\\hline
    $\phi_0$ & $10^{37}\,\mathrm{eV}^{-1}$ & $5.7\pm 1.1\estat$\\
    $\Gamma$ & --- & $2.64\pm 0.08\estat$\\
    \hline
    \multicolumn3c{\textbf{R136 (Hadronic)}}\\\hline
    $\phi_0$ & $10^{37}\,\mathrm{eV}^{-1}$ & $4.0\pm 1.5\estat$\\
    $\Gamma$ & --- & $2.59\pm 0.13\estat$\\
  \enddata
  \tablecomments{
    Statistical errors are at a 68\% confidence level.
    The normalization $\phi_0$ is specified at an energy of $E_0=\SI{3}{TeV}$.
    IC target photon fields assumed in the leptonic models are described in Appendix~\ref{sec:appx_ic_targets}.
    For the fit of the hadronic models, an ambient gas density of \SI{1}{\per\cubic\centi\meter} has been assumed.
  }
\end{deluxetable}

\clearpage
\section{IC Target Photon Fields}
\label{sec:appx_ic_targets}
For the spectral modeling of the \gam-ray emission from \tdorc and R136 with IC models (see Sect.~\ref{sec:results} and Appendix~\ref{sec:appx_naima_results}), we need as input an estimate for the target photon fields.
Besides the cosmic microwave background, we include infrared and optical photon fields that we match to measurements in the 30\,Doradus region taken from \citet{Israel2010} and \citet{Meixner2013}.
These can approximately be described as two blackbody radiation fields with temperatures of \SI{50}{K} and \SI{4000}{K} and energy densities of \SI{2.25}{\electronvolt\per\cubic\centi\meter} and \SI{0.88}{\electronvolt\per\cubic\centi\meter}, respectively.
We note that the estimate for the far infrared field is roughly consistent with that of \citet{HESS_N157B_2012} for the 30\,Doradus region.

In addition, we include for each \tdorc and R136 a dedicated ultraviolet radiation field that describes the emission from the hot, massive stars in the respective star clusters.
For \tdorc, we find from \citet{Testor1993} a mean temperature of \SI{34000}{K} and follow \citet{Haerer2023} to derive an energy density of \SI{10}{\electronvolt\per\cubic\centi\meter}.
In the same manner, using the catalog from \citet{Brands2022}, we obtain an effective temperature of \SI{45000}{K} and an energy density of \SI{20}{\electronvolt\per\cubic\centi\meter} for R136.
Without the ultraviolet fields included, the normalization parameters $\phi_0$ of the leptonic models stated in Table~\ref{tab:naima_parameters} are reduced by about~5\%.
Their presence thus does not affect the conclusions drawn in this work.

\bibliographystyle{aasjournal}
\bibliography{bib_r136}{}

\end{document}